\documentclass[intlimits,twoside,a4paper]{article}

\usepackage{amsmath,amssymb}
\usepackage{graphicx}

\usepackage[T2A]{fontenc}
\usepackage[cp1251]{inputenc}

\usepackage{cmpj2}

\issue{2013}{16}{1}{13606}

\doinumber{10.5488/CMP.16.13606}

\title[DPD study of solvent mediated transitions in pores]%
{Dissipative particle dynamics study of solvent mediated transitions
in pores decorated with tethered polymer brushes in the form of
stripes}

\author[Ja.M.~Ilnytskyi, S.~Soko{\l}owski, T. Patsahan]{Ja.M.~Ilnytskyi\refaddr{a1}, S.~Soko{\l}owski\refaddr{a2}, T. Patsahan\refaddr{a1}}

\addresses{
\addr{a1}
Institute for Condensed Matter Physics of the National
Academy of Sciences of Ukraine, \\
1 Svientsitskii St., 79011 Lviv, Ukraine
\addr{a2}
Department for the Modelling of Physico-Chemical Processes,
Maria Curie-Sk{\l}odowska University, \\ 20031 Lublin, Poland
}

\date{Received October 29, 2012, in final form February 1, 2013}
\authorcopyright{Ja.M.~Ilnytskyi, S.~Soko{\l}owski, T. Patsahan, 2013}

\begin{document}
\maketitle

\begin{abstract}
We study self-assembly of a binary mixture of components A and B
confined in a slit-like pore with the walls modified by the stripes of
tethered brushes made of beads of a sort A. The emphasis is on solvent
mediated transitions between morphologies when the composition of the
mixture varies. For certain limiting cases of the pore geometry we
found that an effective reduction of the dimensionality may lead to a
quasi one- and two-dimensional demixing.  The change of the
environment for the chains upon changing the composition of the
mixture from polymer melt to a good solvent conditions provides
explanation for the mechanism of development of several solvent
mediated morphologies and, in some cases, for switching between
them. We found solvent mediated lamellar, meander and in-lined
cylinder phases. Quantitative analysis of morphology structure is
performed considering brush overlap integrals and gyration tensor
components.

\pacs 62.23.St, 36.20.Ey, 61.20.Ja
\end{abstract}

\section{Introduction}\label{sec1}

In recent years investigations of polymer films on solid surfaces have
become one of the most rapidly growing research area in physics,
chemistry, and material science. The reason for such sustained growth
is due to the availability of a wealth of fundamentally interesting
information in thermodynamics and kinetics, such as long and short
range forces, interfacial interactions, flow, and instability
phenomena \cite{1,2,3,4,5,6,7,9}. Moreover, polymer thin films are
widely used as an industrial commodity in coatings and lubricants and
 have become an integral part of the development process in modern
hi-tech applications such as optoelectronics, biotechnology,
nanolithography, novel sensors and actuators \cite{a1,a2,a3,a4,a5,a6,a7,a8}.
Most of these applications have been
connected with an intrinsic property of polymer films to exhibit a
variety of surface morphologies, the size of which ranges from a
few tens of nanometers to hundreds of micrometers.  Since many of the
modern technologically relevant phenomena occur at the nanoscale, the
behavior of polymer thin films deposited on a flat surface that exhibit
morphologies at the nanoscale has been one of major focuses in recent
years \cite{6,9,r1,r2,r3,r4,r5,r6}.

The development of several new techniques \cite{nt} in material science
permits now the productions of solid substrates whose surface is
``decorated'' with precisely characterized surface structures on
the length scales ranging from nanometers to microns.  In particular,
advances in nanotechnology have permitted the establishment of methods for
obtaining functional polymeric films on solid surfaces exhibiting
quite complex topographic nanostructures.  Such chemically decorated
substrates allow for manipulation of fluid at very short length scales
and can play an important role in a variety of
contexts \cite{th1,lito,lito_1,lito_2,nano,nano_1,nano_2}.

The importance of systems involving brushes tethered at structured
surfaces urged the development of methods for theoretical description
of such systems.  Theoretical studies of fluids in contact with
brushes on patterned surfaces have been mainly based on different
simulation methods. They include Monte Carlo \cite{mc,mc_1,mc_2,mc_3,mc_4} and molecular
dynamics \cite{md,md_1}, as well as dissipative particle dynamics (DPD)
simulations \cite{r1,dpd,dpd_1,dpd_2,dpd_3,dpd_4,dpd_5}. The studies performed so far indicated that
heterogeneity of tethered layers has a great impact on the structure
of the confined fluid and on thermodynamic and dynamic properties of the
systems.

In numerous previous studies \cite{dpd,dpd_1,dpd_2,dpd_3,dpd_4,dpd_5}, including our work \cite{ISP1}, the
simulations have been carried out assuming constant composition
in the confined system. However, the demixing phenomena and the
formation of different morphologies strongly depend on
 the composition \cite{IL}. Thus, the aim of this work is to determine the effect of the change
in the fluid composition on the structure of a
confined system. Similarly to the previous work \cite{ISP1},
 we use DPD to investigate the
behavior of a binary mixture composed of beads A and B confined in
slit-like pores with walls modified by the stripes of tethered chains made of beads A. The stripes at the opposing pore walls are
placed  ``in-phase'' (face-to-face), or ``out-of-phase''.
Species A and B are assumed to exhibit a demixing
in a bulk phase.  Simulations are carried out for different
compositions of the fluid.
 Our interest is to determine possible morphologies that
can be formed inside the pore depending on the fluid
composition as well as on the geometrical
parameters characterizing the system (the size of the pore and the
width of the stripes, the arrangement of the stripes).
In particular, we analyze special limiting cases, where geometry of the pore
leads to the reduction of the effective dimensionality. Special
emphasis is made on the cases of stripes between which the separation distance, either within the pore wall or across the pore, is
small. The crossover for the polymer chains from the regime of polymer
melt to the regime of a good solution provides a basis for solvent
mediated morphology formation and morphology switching. We performed
complementary simulations where the solvent in a form of a binary
mixture is replaced by a one-component solvent of variable
quality. Moreover, we also consider
how the arrangement of the stripes (``in-'' versus ``out-of-phase'')
effects the observed phenomena.
Quantitative analysis of morphologies is
performed by means of overlap integrals of polymer chains belonging to
different stripes or surfaces. Conformational properties of chains are
studied via gyration tensor components. The simulation method was
described in our previous work \cite{ISP1} and for the sake of brevity
is omitted here.

The paper is organized as follows. In the next section we will briefly recall
 the model and the simulation method. A number of limiting
cases for the pore geometry that show the effects of a reduced effective
dimensionality are considered in section~\ref{sec3}. Quantitative
analysis of solvent mediated morphologies for the case of weakly
separated stripes is performed in section~\ref{sec4}. The summary of
the results is presented in section~\ref{sec5}.

\section{The model} \label{sec2}

\begin{figure}[!t]
\centering \includegraphics[width=10cm]{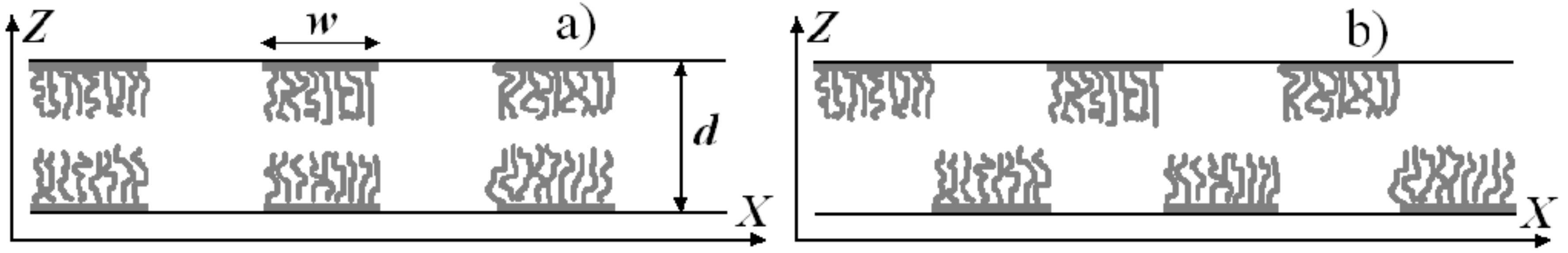}
\caption{\label{Fig-01r}Geometry of the pore with in-phase (a) and
  out-of-phase (b) arrangement of the stripes with tethered polymer
  chains. The pore size is $d$ and the stripes width is $w$. The
  length of polymer chains is $L=20$ beads.}
\end{figure}

To simulate the pore, we use the box of the dimensions
$L_x$, $L_y$ and $L_z$. All the dimensions are in reduced units
measured with respect to the cutoff distance $r_{\mathrm{c}}$ for the repulsive
interaction, which is set to $r_{\mathrm{c}}=1$.  The planes $z=0$ and
$z=L_z\equiv d$ are impenetrable walls.  Periodic boundary conditions
are applied in both $X$ and $Y$ directions. Each wall is divided into
stripes of equal width, $w$, alternating those with and without
polymers attached.  The stripes at the opposite walls can be placed
either in- or out-of-phase, see figure~\ref{Fig-01r}. The width of the
stripes with attached polymers is the same as the width of the polymer
free spaces. Therefore, the term ``out-of-phase'' means that the stripe
with polymer on one wall  faces the polymer free stripe at the other
wall.  The rest of the pore interior is filled with single beads
representing the fluid components.

We use the DPD method in the form outlined by Groot and Warren
\cite{GrWarr}. This is a mesoscopic approach in which one considers
beads that interact via soft repulsive potential. Each bead can be
interpreted as a fragment of a polymer chain or a collection of
solvent particles, etc. The strength of the repulsive interaction can
be tuned to match the compressibility of e.g. water, whereas the
difference in repulsion for various species reflects their miscibility
and can be related to the Flory-Huggins parameter \cite{GrWarr}.  The
pairwise dissipative (friction) and random forces are added in a
spirit of Langevin dynamics to compensate for averaging over
a small-scale structure being neglected in this approach. The method
proved itself to be fast and effective for the study of
microphase-separation driven phenomena \cite{dpd,dpd_1,dpd_2,dpd_3,dpd_4,dpd_5}.

The method is extended to include the smooth repulsive walls via the
reflection algorithm that preserves the total momentum
\cite{ISP1}. The reduced number density for beads is equal to
$\rho=3.0$.  Polymer chains are built of $L=20$ beads linked via
harmonic bonds.  The grafting points have been fixed on each wall and are
distributed randomly inside the stripes with a reduced grafting density
$\rho_\mathrm{g}=1.0$. More details are provided elsewhere \cite{ISP1}.

Two bead sorts, A and B, are considered. Polymer chains are made of
beads of sort A.  The total number of polymer beads in the system is
$N^{\mathrm{p}}_{\mathrm{A}}$. A binary fluid mixture (which fills the interior of the pore)
contains $N^{\mathrm{s}}_{\mathrm{A}}$ beads of sort A and $N_{\mathrm{B}}$ beads of sort B.  The
fraction of the beads of sort B is $f_{\mathrm{B}}=N_{\mathrm{B}}/N$, where
$N=N^{\mathrm{p}}_{\mathrm{A}}+N^{\mathrm{s}}_{\mathrm{A}}+N_{\mathrm{B}}$ is the total number of the beads.
Repulsion parameters for A--A and B--B pairs are defined to be the same,
$a_{\mathrm{AA}}=a_{\mathrm{BB}}=25$, the choice being based on matching the
compressibility of the model system (at $\rho=3.0$) to that of water,
as discussed in \cite{GrWarr}. Poor miscibility between A and B
beads is set by the higher value of $a_{{\mathrm{AB}}}$ for the A--B interaction
as compared to $a_{{\mathrm{AA}}}$. The choice of $a_{{\mathrm{AB}}}$ value is quite
flexible and we use the value of $a_{{\mathrm{AB}}}=40$ which corresponds to the
regime of moderate to strong segregation and ensures firm separation
between species.

To visualize the  morphologies, we employ the following density grid
approach.  Simulation box is split into the grid of cubic cells with
the linear dimension of $0.75\div2.0$, depending on the simulation box
size. Local densities for polymer A beads, $\rho^{\mathrm{p}}_{\mathrm{A}}(x,y,z)$, solvent
A beads, $\rho^{\mathrm{s}}_{\mathrm{A}}(x,y,z)$, and solvent B beads, $\rho_{\mathrm{B}}(x,y,z)$ are
evaluated in each cell centered at $(x,y,z)$. They are averaged over
$10-15$ configurations (the subsequent configurations are separated by
$5000$ simulation steps). The averaging is carried out after
the particular  morphology is stabilized (typically, after $2\cdot 10^5$
simulation steps).

We found that presenting both local densities of A and B beads in the
same snapshot is not very informative.  Instead, we show
separate snapshots for local density of A or of B
beads. In the first case, we represent the cells with low local
density of A beads $[\rho^{\mathrm{p}}_{\mathrm{A}}(x,y,z) +\rho^{\mathrm{s}}_{\mathrm{A}}(x,y,z)] /\rho<0.15$ as
dots. All other cells are space-filled with the color
saturation proportional to the value of $[\rho^{\mathrm{p}}_{\mathrm{A}}(x,y,z)
+\rho^{\mathrm{s}}_{\mathrm{A}}(x,y,z)] /\rho$. The color tint is blueish if
$\rho^{\mathrm{p}}_{\mathrm{A}}(x,y,z) >\rho^{\mathrm{s}}_{\mathrm{A}}(x,y,z)$ and greenish otherwise.  Similar
approach is used in the second case, when the density of B beads is
plotted. In this case, $\rho_{\mathrm{B}}(x,y,z)$ is color coded with red tint.

\section{Solvent-mediated transitions in special geometries}\label{sec3}

\subsection{Wide stripes geometry}

\begin{figure}[!t]
\centering \includegraphics[width=8cm]{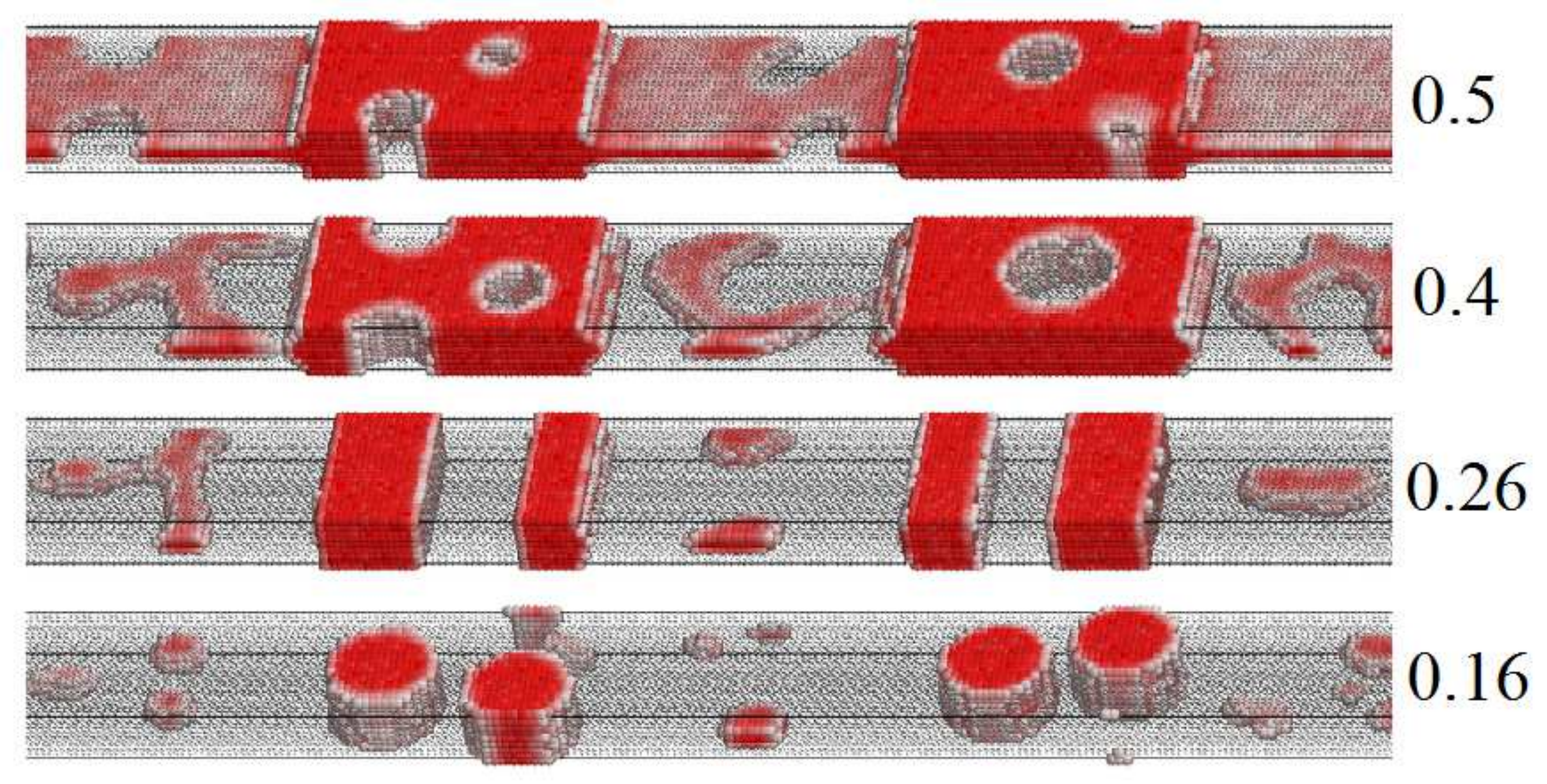}
\caption{(Color online) \label{Fig-02r}The sequence of morphologies obtained for the
  case of wide stripes, $w=90$, at pore size $d=20$ by varying
  fraction of B beads, $f_{\mathrm{B}}$ (indicated on the right). Color-coded
  density of B beads in the selected part of the pore is shown only
  for the sake of clarity.}
\end{figure}

In the case of wide stripes that are in-phase arranged, the periodic
pattern of alternating sub-regions is formed within the pore. The
sub-regions free of polymer brush provide an environment for a bulk,
quasi-3D demixing of confined A and B beads. On the contrary, the
sub-regions dominated by a brush, reduce the volume accessible for the
demixing of A and B beads to quasi-2D slabs, especially for moderate
values of pore size $d$.  This is demonstrated in figure~\ref{Fig-02r},
where we show examples of the morphologies observed for wide stripes with
$w=90$ in a pore of size of $d=20$ when the fraction $f_{\mathrm{B}}$ is varied.

The first effect to mention is that with an increase of the fraction
of A beads, $f_{\mathrm{A}}=1-f_{\mathrm{B}}$, more of them are adsorbed by the stripes of
polymer brush causing their considerable swelling. We will discuss
this effect quantitatively in section~\ref{sec4}. The remaining A and
B beads are spread within the rest of the accessible volume.

We will concentrate here on the phenomena that occur within quasi-3D
sub-regions. The morphologies formed here repeat those observed for
diblock copolymers at various composition (or in similar
systems) \cite{ISP1,IL}. To relate both cases, one should look at the
local fractions of A and B beads, $f'_{\mathrm{A}}$ and $f'_{\mathrm{B}}$, that differ
from their global counterparts, $f_{\mathrm{A}}$ and $f_{\mathrm{B}}$, due to adsorption of
some of A beads into the brush mentioned above. We found that
cylinders made of B beads are formed at $f_{\mathrm{B}}=0.16$ ($f'_{\mathrm{B}}=0.31$,
lowest frame in figure~\ref{Fig-02r}). Similarly, cylinders made of A
beads are observed inside the interval from $f_{\mathrm{B}}=0.40$ ($f'_{\mathrm{A}}=0.26$)
to $f_{\mathrm{B}}=0.50$ ($f'_{\mathrm{A}}=0.17$), first and second frames from the top of figure~\ref{Fig-02r}. Lamellar-like phase (in the $OYZ$ plane;
alternating blocks of A and B beads along the $X$-axis) is observed in
the interval from $f_{\mathrm{B}}=0.26$ ($f'_{\mathrm{A}}=0.50$) to $f_{\mathrm{B}}=0.30$
($f'_{\mathrm{A}}=0.42$), second frame from the bottom in figure~\ref{Fig-02r}. These boundaries of morphologies (in terms of the
values $f'_U$ for minor fraction $U$) correlate well with the phase
diagram for diblock copolymers (see, e.g., \cite{ISP1,IL}).

Morphology transformations observed within the quasi-2D regions literally
repeat those found for the narrow stripes geometry, and this case is
considered in detail in the following subsection.

\subsection{Narrow stripes geometry}\label{subs_narrow}

For narrow stripes, $w<5$, the polymer chains from adjacent stripes
bridge themselves into lamellae that envelope each wall \cite{ISP1}.
It is quite obvious that the same scenario holds for the out-of-phase
arrangement of stripes, as far as both surfaces are decoupled. This is
confirmed in our simulations (not shown for the sake of brevity). Two
flat lamellae, formed at each wall, reduce the region accessible to
free A and B beads to a slab, which turns into a quasi-2D one for
moderate pore size $d \sim 14\div20$.  This situation also represents the
sub-regions dominated by a brush for the case of wide stripes
(considered in the previous subsection).

\begin{figure}[!b]
\centering \includegraphics[width=12cm]{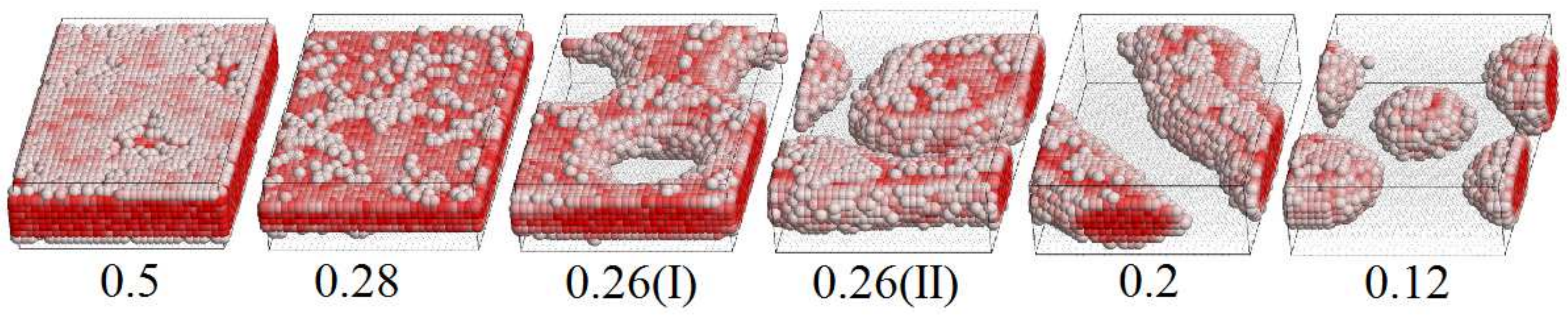}
\caption{(Color online) \label{Fig-03r}Sequence of morphologies obtained for narrow
  stripes, $w=4$, in a pore of size $d=13.333$ at various fractions
  $f_{\mathrm{B}}$ (indicated at the bottom). From left to right: lamellar,
  thinned lamellar, perforated lamellar, ``sausage'' and ``cake'',
  single ``sausage'', and hexagonally packed ``cakes'' morphologies
  are shown (only color-coded density of B beads is shown).}
\end{figure}

In figure~\ref{Fig-03r}, we display the sequence of quasi-2D morphologies
that appear in this central slab as the result of a micro-phase
separation between A and B beads. The pore size is $d=13.333$, the
stripes width is $w=4$ and only color-coded density of B beads is
shown (in red tint). This pore size is a special one, as far as for
the parameters being used here (polymer length, bulk and grafting
densities), no solvent A beads are present at $f_{\mathrm{B}}=0.5$ \cite{ISP1}. At
this fraction $f_{\mathrm{B}}$, solvent B beads fill-in all the slab-like accessible
volume in the middle of the pore (first frame from the left in
figure~\ref{Fig-03r}). The slab stays continuous but thins out with a
decrease of $f_{\mathrm{B}}$ down to $f_{\mathrm{B}}=0.28$ (the next frame). Then, it turns
into a perforated lamellar one and then into disjointed prolate and/or
oblate objects made of beads B (see respective frames in
figure~\ref{Fig-03r}). For still lower value of $f_{\mathrm{B}}$, $f_{\mathrm{B}}\approx 0.12$, there appear
hexagonally distributed ``cookies'' of B beads. One can
easily see that this morphology possesses the same symmetry as
the perforated lamellar one, if beads A and B are interchanged.  Further
decrease of $f_{\mathrm{B}}$ causes the loss of  hexagonal order of the ``cookies'', and for $f_{\mathrm{B}}<0.1$ the ``cookies'' of B beads become randomly
arranged (not shown). This sequence of the morphologies is reminiscent
of the one observed in partial mixing of the two-dimensional
fluids \cite{Zhao}. Therefore, the case of narrow stripes can be
classified as geometry-driven dimensional crossover from 3D to 2D for
a confined fluid.

\subsection{Pillar geometry}\label{subs_pillar}

\begin{figure}[!t]
\centering \includegraphics[width=13cm]{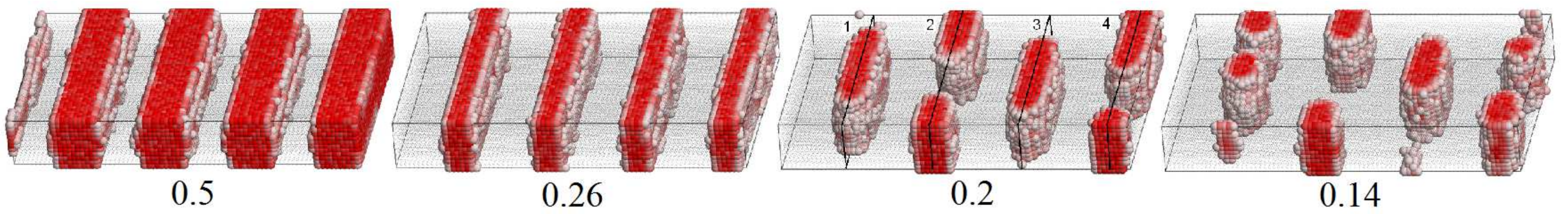}
\caption{(Color online) \label{Fig-04r}Sequence of morphologies obtained for
  $d=13.333$ and $w=10$ at various fractions $f_{\mathrm{B}}$ (indicated at the
  bottom). Blocks of B beads filling the space between pillars of A
  beads ($f_{\mathrm{B}}=0.50$), thinned blocks ($f_{\mathrm{B}}=0.26$), hexagonally
  arranged columns ($f_{\mathrm{B}}=0.20$), and random columns ($f_{\mathrm{B}}=0.14$) are
  shown. Only color-coded density of B beads is displayed.}
\end{figure}

\begin{figure}[!b]
\centering \includegraphics[width=8cm]{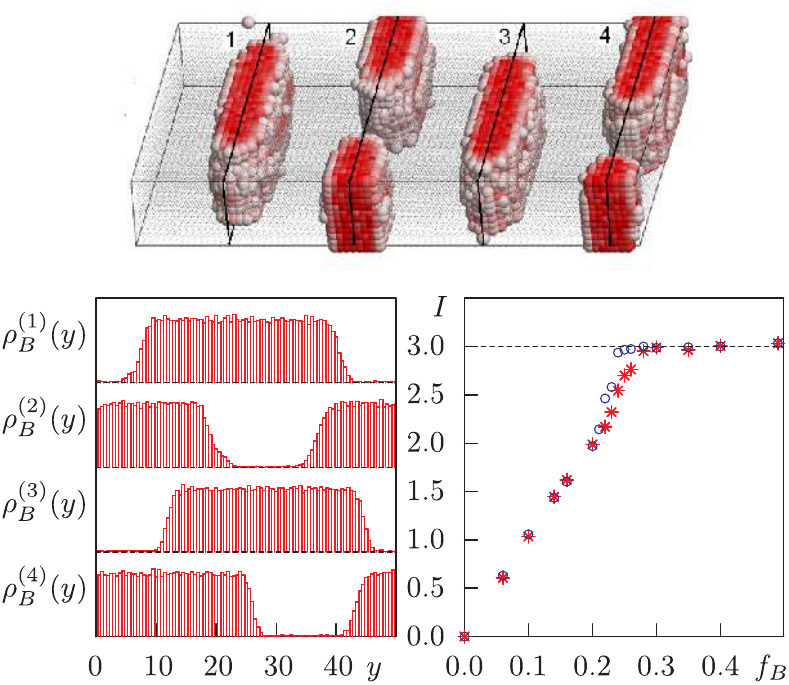}
\caption{(Color online) \label{Fig-05r}Left bottom frame shows the density profiles  $\rho^{(i)}(y)$ along each $i$-th slab region shown in the top frame.
Right bottom frame shows  the dependence of the integral profile $I$ (averaged over four  slabs) on $f_B$.
The results for simulation set I are shown as blue  disks and for the set II~--- as red asterisks.}
\end{figure}

In the case of wider stripes ($w>6$), bridging of polymer chains that
belong to adjacent stripes is prohibited due to high penalty in
conformational entropy. Instead, the in-phase arranged stripes can
bridge themselves across the pore to form pillars, providing that the
pore width $d$ is not too big (cf. the sketch phase diagram presented
in the previous work \cite{ISP1}). Such pillars are observed, for
example, at $d=13.333$ and $w=10$. Similarly to the case shown in
figure~\ref{Fig-02r}, the system again displays (periodic along $X$
axis) the pattern of sub-regions, one being pillars of merged brushes and
another being polymer-free sub-regions. At $f_{\mathrm{B}}=0.5$, the latter are in a
form of blocks filled exclusively with solvent B beads. For
$0.5>f_{\mathrm{B}}>0.26$, these blocks become thinner first (as far as solvent A
beads are adsorbed into pillars causing the latter to swell).  Then,
for $f_{\mathrm{B}}\sim 0.20$, the blocks split into rounded ``columns'' spanning
across the pore and arranged almost hexagonally.  At still lower
values of $f_{\mathrm{B}}$, we arrive at randomly arranged columns of B beads of
random thickness.  All these morphologies are displayed in
figure~\ref{Fig-04r}.

The regions accessible to the micro-phase separation of solvent beads
have a shape of slabs extended along $Y$-axis. The important point
here is that all morphologies observed within these slabs at various
$f_{\mathrm{B}}$ are uniform within the $X$ dimension of each slab. Thus, the
behavior of the system can be interpreted as being quasi-1D along $Y$
axis within each slab. One can quantify the observed morphology changes by
the density profiles of B beads, $\rho(y)$, along $Y$ axis. It can be
evaluated within a thin cross-section slabs located in the middle of
each region (marked as $1-4$ in the top frame in
figure~\ref{Fig-05r}). The thickness of each cross-section slab in $X$
direction is equal to $2$ and the averaging of the density is made in
$Z$-direction. The histograms for the density profiles $\rho^{(i)}(y)$
in each $i$-th slab obtained in this way for the morphology shown in the top frame in figure~\ref{Fig-05r}, are presented in the left bottom frame of figure~\ref{Fig-05r}. The local density inside each
column of B beads is equal to the bulk density,
$\rho(y)\approx\rho=3$, whereas it drops down to
zero outside the column.  Therefore, the integrated density, $I=\int_{0}^{L_y}
\rd y\rho(y)/L_y$ is a good measure for ``block continuity''.  Right hand
frame in figure~\ref{Fig-05r} shows the values of $I$ averaged over all
four slabs, $1-4$. Two sets of simulations were performed. For the set
I we started the simulation from the morphology equilibrated at
$f_{\mathrm{B}}=0.5$ and converted the required number of B beads, chosen randomly
across the system, into A beads.  In the set II simulation, the
initial configuration involved linearly stretched polymer chains and
solvent A and B beads randomly distributed within the pore.

One can observe the transformation from continuous into discontinuous
block morphology that occurs at $f^*_{\mathrm{B}}\approx 0.23$ (for the
simulations set I; blue open disks) or at $f^*_{\mathrm{B}}\approx 0.28$ (for the
simulation set II; red asterisks). The former value is lower
which indicates the presence of the ``underconcentrated'' continuous block
morphology at $0.28>f_{\mathrm{B}}>0.23$ in the simulations set I. It is also
interesting to note that for small $f_{\mathrm{B}}$, the integral $I$ changes
nearly linearly with $f_{\mathrm{B}}$.

\section{Competition between solvent mediated morphologies for closely\\
arranged stripes}\label{sec4}

So far we considered several special cases when both the geometry
restrictions imposed by the pore geometry and the composition of the
mixture confined within the accessible sub-regions promote essential
morphological changes. In both cases of a narrow pore (subsection~\ref{subs_narrow}) and of a pillar (subsection~\ref{subs_pillar}), the
stripes of brushes are bridged in one of the directions by purely geometrical
means, literally by bringing stripes close enough to form lamellae
(pillars). The role of the solvent is restricted to the swelling of the
already formed lamellae (pillars) and to a micro-phase separation inside
the polymer-free sub-regions.

However, one can envisage the situation when the lamellar (pillar)
bridges are formed exclusively due to the action of a solvent.
Apparently, this is possible in the geometries where the stripes of brush
are brought sufficiently close in one of $X$ or $Z$ dimensions, but
not close enough to bridge over by themselves. Even a more promising
case can be designed by provoking the competition between the bridging,
which may happen when the stripes are positioned sufficiently close in
both $X$ and $Z$ directions. Possible technological applications could
involve the use of a solvent mixture which contains A component in the
form of a short chain. This component could be first used for demixing
with the B component and for the formation of certain morphology, and then to be
used as a crosslinker to permanently fix the structure.

\subsection{Mechanism for solvent-mediated morphologies and
 quantitative characterization of brush bridging and deformation of  chains}

The mechanism for the solvent-mediated morphologies lies in the swelling
of a polymer brush due to adsorption of a good solvent, the effect
already mentioned above. Here, we will discuss this effect in more
detail. Let us consider the environment within the stripes of polymer
chains. Due to relatively low grafting density
$\rho_\mathrm{g}=\frac{1}{3}\rho$ (where $\rho=3$ is bulk number density), the
stripes are far from the regime of a dense brush \cite {Romeis_2012} and
are exposed to the available solvent. For the case of poor solvent, the
brush collapses and each chain is in a regime of polymer
melt, whereas for a good solvent, the chains are expected to be in the
regime of a good solution. Two cases are well distinguished by their
respective scaling laws~\cite{deGennes}. For instance, the radius of
gyration scales as:
\begin{equation}\label{scaling}
 R_\mathrm{g} = R \left[l_0 (N-1)\right]^\nu\,,
\end{equation}
where $R=\mathrm{const}$, $l_0$ is the equilibrium bond length, $N$
is the number of monomers and $\nu=0.5$ for the case of polymer melt
and $\nu\approx 0.59$ (Flory exponent) for the case of a good
solution \cite{deGennes}. As it was discussed previously \cite{IlnHol},
the softness of the potentials employed in typical DPD simulations
does not violate the correct value for the exponent $\nu=0.59$ for a
single chain in a good solvent.

To check whether this scenario holds, we performed a set of
simulations for the pore geometry with well separated stripes.
Polymer chains are made of A beads and a one-component solvent of C
beads is used. The quality of the solvent is tuned via the repulsion
parameter $a_{{\mathrm{AC}}}$ between A and C beads ranging from $25$ (good
solvent) up to $40$ (bad solvent). For each simulation run, the
average bond length $l_0$ and the radius of gyration $R_\mathrm{g}$ were evaluated
and the scaling law (\ref{scaling}) was employed for the assumed
values of the exponent $\nu$, $\nu=0.5$ or $0.59$. Next, the prefactor
$R$ (see equation~(\ref{scaling}) was estimated. The results are collected
in table~\ref{table_scaling}. As it follows from the enclosed data,
the difference between the estimated prefactors $R$ does not exceed
1\% which proves that the scaling law (\ref{scaling}) holds
consistently. These simulations confirm a crossover from the regime of
polymer melt to a good solution regime for each chain which is driven
by the quality of the solvent. Typical chain extension ratio is
estimated as $[l_0 (N-1)]^{0.09}\sim 1.3$ which is of the order of
$2\div4$ unit lengths for our particular model, and this provides the
length scale for brush separations at which one expects
solvent-mediated bridging.
\begin{table}[!t]
\caption{Results of fitting the average bond length $l_0$ and radius
  of gyration $R_\mathrm{g}$ to the scaling law, equation~\protect\ref{scaling}, for
  the assumed values of the exponent $\nu$. Output: prefactor $R$ in
  equation~\protect\ref{scaling} is numerically consistent for all the cases
  considered. \label{table_scaling} }
  \vspace{2ex}
\centering
\begin{tabular}{|c|c|c|c|c|}
\hline\hline
pore geometry         & solvent  & $l_0$ & $R_g$ & $R$ \\
\hline\hline
$d=22$, in-phase   & $a_{{\mathrm{AC}}}=40$ $(\nu=0.50)$  & 0.893 & 1.909 & 0.463\\
   & $a_{\mathrm{AC}}=25$ $(\nu=0.59)$  & 0.937 & 2.557 & 0.468\\
\hline
$d=20$, out-of-phase & $a_{\mathrm{AC}}=40$ $(\nu=0.50)$ & 0.892 & 1.909 & 0.464\\
  & $a_{\mathrm{AC}}=25$ $(\nu=0.59)$ & 0.936 & 2.558 & 0.468\\
\hline\hline
\end{tabular}
\end{table}

\begin{figure}[!t]
\centering \includegraphics[width=0.95\textwidth]{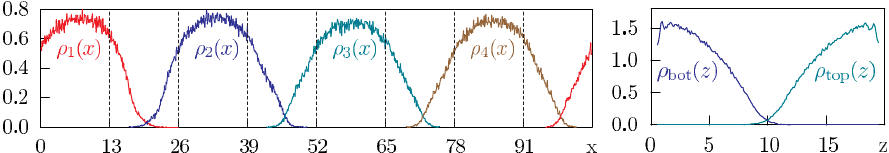}      
\caption{(Color online) \label{Fig-06r}Left hand frame illustrates the density distributions
  $\rho_i(x)$ of the polymer beads in each $i$-th stripe along $X$
  axis averaged over $Y$ and $Z$ axes (the distributions $1-4$
  belonging to the bottom surface are shown only). Right hand frame
  illustrates the density distributions of the polymer beads along $Z$
  axis that belong to bottom and top surfaces (abbreviated as ``bot''
  and ``top'', respectively), the distributions are averaged in $X$ and
  $Y$ directions.}
\end{figure}

To quantitatively characterize the level of bridging between stripes
we introduce the overlap integrals $I_x$ and $I_z$ between polymer
chains in $X$ and $Z$ directions (by analogy to the case of uniform
brushes \cite{Kreer_2001}):
\begin{equation}\label{integral_xz}
I_x = \sum_{\langle ik \rangle}\int_{0}^{L_x}\rho_i(x)\rho_k(x) \rd x
\cdot\left[\sum_{\langle ik \rangle}\int_{0}^{L_x}\rd x\right]^{-1} \qquad
I_z = \int_{0}^{d} \rho_{\mathrm{bot}}(z)\rho_{\mathrm{top}}(z) \rd z
\cdot\left[\int_{0}^{d} \rd z\right]^{-1}\!\!\!\!\!\!\!\!,
\end{equation}
where $\rho_i(x)$ is the density profile along $X$ axis for polymer
beads belonging to $i$-th stripe (averaged over $Y$ and $Z$
directions); $\rho_{\mathrm{bot}}(z)$ and $\rho_{\mathrm{top}}(z)$ are
density profiles for polymer beads belonging to the bottom or top wall,
respectively (averaged over $X$ and $Y$ directions). These density
profiles are illustrated in figure~\ref{Fig-06r}. The evaluation of
$I_x$ overlap integral is performed over all $\langle ik \rangle$
pairs that are located at the same wall and are adjacent in $X$
direction (with the account of the periodic boundary conditions).

The bridging of brushes involves a certain amount of bending and/or
anisotropic stretching of polymer chains. These can be quantified via
the components of gyration tensor, $G_{xx}$, $G_{yy}$ and $G_{zz}$.
The components are evaluated for each $k$-th chain
\begin{equation}\label{Gtens}
G_{\alpha\beta}^{[k]} = \frac{1}{N}\sum_{i=1}^{N}
   \left(r_{i,\alpha}^{[k]}-R_\alpha^{[k]}\right)\left(r_{i,\beta}^{[k]}-R_\beta^{[k]}\right),
\end{equation}
where $\alpha,\beta$ denote Cartesian axes, $r_{i,\beta}^{[k]}$ are
the positions of individual monomers and $\mathversion{bold}{R}^{[k]}$ is the
center of mass position of the $k$-th chain. Then,
$G_{\alpha\beta}^{[k]}$ are averaged over the chains and over the time
trajectory after the morphology stabilizes itself, providing the
estimates for $G_{xx}$, $G_{yy}$ and $G_{zz}$.  One should mention
that due to the symmetry of the pore in $Y$ direction, the $G_{yy}$
component is found to be unchanged and it is not considered in our
analysis. The average radius of gyration $R_\mathrm{g}$ is defined as
$R_\mathrm{g}^2=G_{xx}+G_{yy}+G_{zz}$, and the average extension of chains can
be found from the maximal eigenvalue of the gyration tensor,
$\sigma_{\max}^2$.

\subsection{Solvent-mediated morphological changes}

\begin{figure}[!b]
\centering \includegraphics[width=8cm]{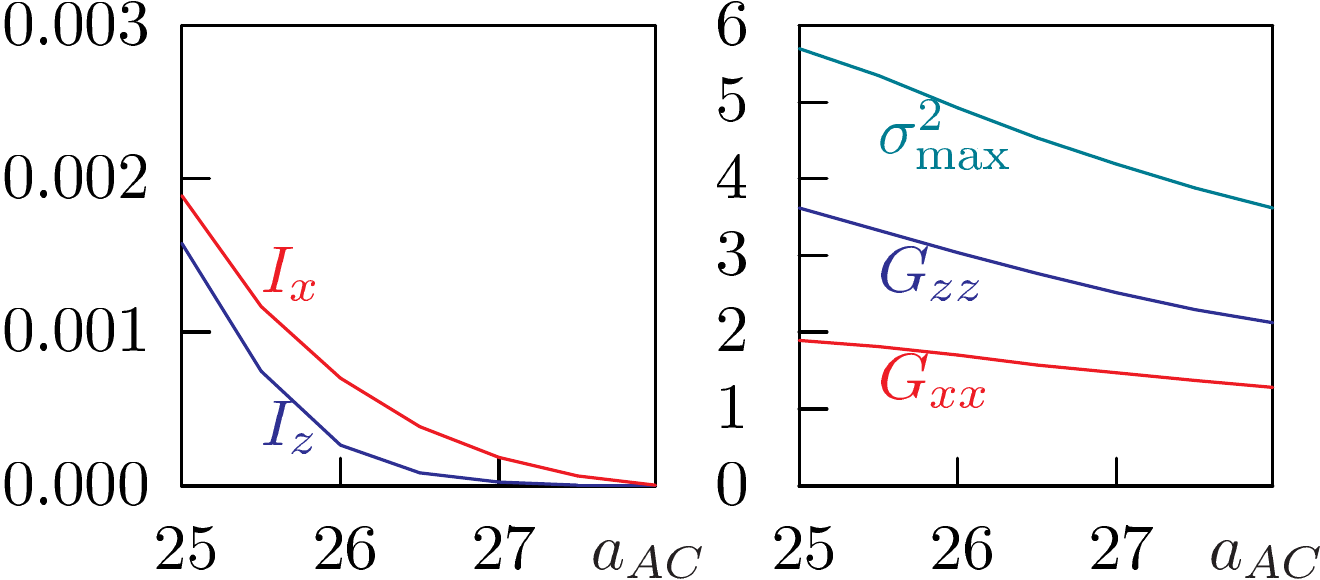}
\caption{(Color online) \label{Fig-07r}Behavior of the overlap integrals $I_x$ and
$I_z$ (left hand frame) and the components $G_{xx}$ and $G_{zz}$ and
maximal eigenvalue $\sigma_{\max}^2$ of the gyration tensor (right hand frame)
upon the changes of the quality of the one-component solvent. The
quality of the solvent is defined via the repulsion parameter $a_{\mathrm{AC}}$
between polymer (A) and solvent (C) beads, the value of $a_{\mathrm{AC}}=25$
represents the case of a good solvent.
}
\end{figure}

The properties introduced above are used to characterize
solvent-mediated bridging between stripes of brushes. At first we will
consider the case of one-component solvent of variable quality (tuned
via $a_{\mathrm{AC}}$ parameter, see above). The results are provided for the
pore size of $d=20$ and stripes of width $w=13$ arranged out-of-phase
(see figure~\ref{Fig-07r}).  With a decrease of $a_{\mathrm{AC}}$ towards the case
of a good solvent ($a_{\mathrm{AC}}=25$), one observes a monotonous increase
of all the metric properties, $G_{xx}$, $G_{zz}$ and
$\sigma_{\max}^2$. This quantifies the effect of polymer chains
swelling due to the crossover from the polymer melt to polymer in a
good solvent regime. The overlap integrals $I_x$ and $I_z$ have been
found to become non-zero at a certain threshold value of
$a_{\mathrm{AC}}<28$. This indicates a solvent-mediated bridging between the droplets
of polymer chains. The bridging happens almost simultaneously in both
$X$ and $Z$ directions for this geometry of the pore. In view of
our analysis, it is important to mention that all the characteristics,
$I_x$, $I_z$, $G_{xx}$, $G_{zz}$ and $\sigma_{\max}^2$ increase
monotonously with a decrease of $a_{\mathrm{AC}}$ down to $25$ with no
peculiarities observed either in the behavior of metric properties
and overlap integrals or in the snapshots.

The case of a one-component solvent of the variable quality provides a
suitable reference point for a more complex case of a two-component
solvent composed of A and B beads. The repulsion parameter $a_{\mathrm{AB}}$ is
chosen to be always equal to $40$ in our study. Therefore, the
effective ``goodness'' of such a mixture can be characterized by the
fraction $f_{\mathrm{B}}$ of B beads (the mixture is a good solvent for
$f_{\mathrm{B}}=0$). The analogy with a one-component solvent is, however, not
exact, as far as the mixture is prone to segregation (see, e.g. figure~4
in \cite{ISP1}). Local inhomogeneities may effect the way in
which the brushes are bridged and, as a result, the sequence of
solvent-mediated morphologies may differ from the case of
a one-component solvent. Hereafter, we will consider the fixed stripe width
of $w=13$ and both: in- and out-of phase arrangements. The pore size
will be fine-tuned in each case and ranges from $d=19$ to $d=24$.

\begin{figure}[!t]
\centering \includegraphics[width=10cm]{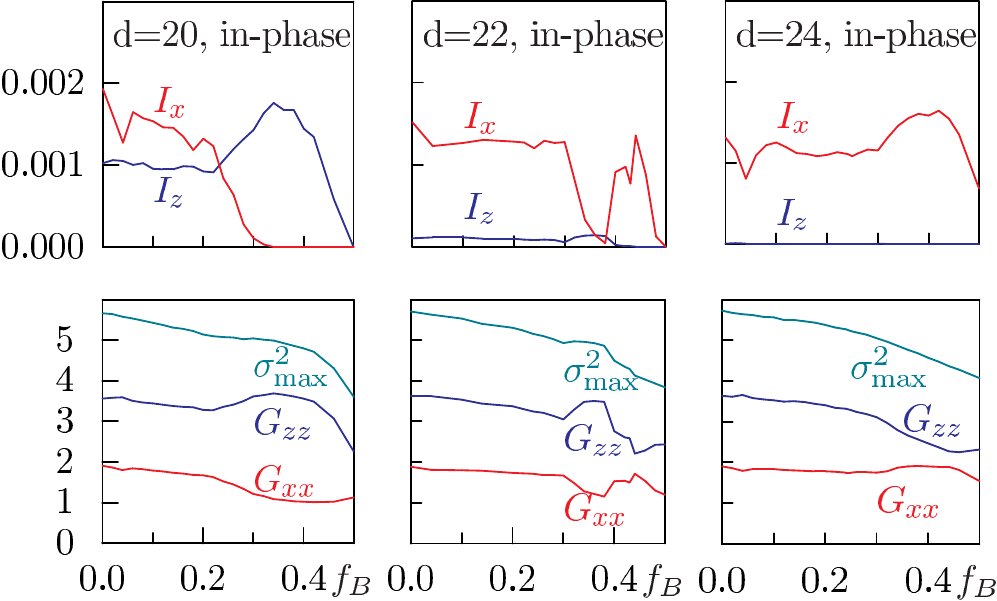}
\caption{(Color online) \label{Fig-08r}Behavior of the overlap integrals $I_x$ and
  $I_z$ (top row of frames) and the components $G_{xx}$ and $G_{zz}$
  and the maximal eigenvalue $\sigma_{\max}^2$ of the gyration tensor
  (bottom row of frames) upon the changes of the effective quality of
  the two-component solvent of A and B beads. Effective quality of the
  solvent is defined via the fraction $f_{\mathrm{B}}$ of ``bad solvent'' beads
  B. The case of in-phase arrangement of stripes of the width $w=13$
  is shown at the pore sizes $d=20, 22$ and $24$.  }
\end{figure}

In the case of in-phase arrangement of the stripes, we select three
pore sizes, namely $d=20, 22$ and $24$ (the distance between brushes
in $X$ and $Z$ directions is approximately the same for $d=22$). The
evolution of $I_x$, $I_z$, $G_{xx}$, $G_{zz}$ and $\sigma_{\max}^2$
upon the change of the fraction of $f_{\mathrm{B}}$ from 0.5 down to 0 is shown
for each pore size in figure~\ref{Fig-08r}. One can compare these plots
with the curves displayed in figure~\ref{Fig-07r} and observe that the
metric properties of chains at $f=0.5$ subject to the two-component
solvent, match approximately their counterparts for the case of
a one-component solvent at $a_{\mathrm{AC}}=28$. This provides the estimate for
an effective ``goodness'' of the two-component solvent. Metric
properties coincide for the case of good solvent ($f=0$ and
$a_{\mathrm{AC}}=25$, respectively).

\begin{figure}[!b]
\centering \includegraphics[width=12.2cm]{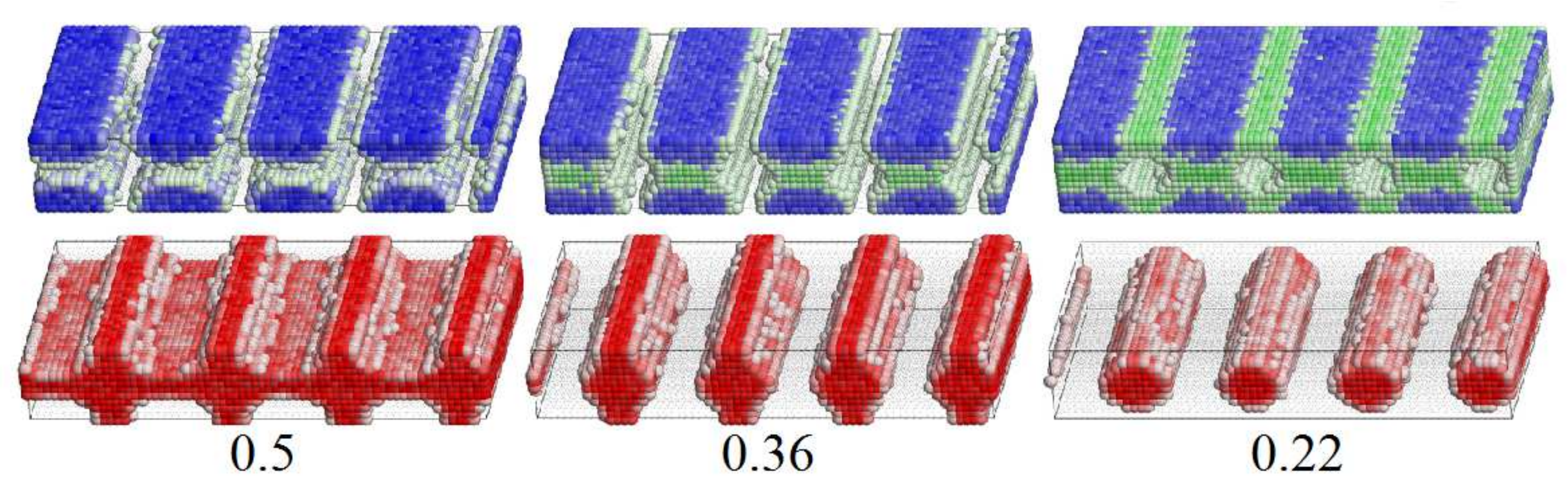}
\caption{(Color online) \label{Fig-09r}Sequence of morphologies obtained for $d=20$
  and for in-phase arranged stripes of the width $w=13$ at various
  fractions $f_{\mathrm{B}}$ (indicated at the bottom). Top row: color-coded
  densities of brush prevailing (blue) and good solvent prevailing
  (green) regions, bottom row: bad solvent prevailing regions (red).
  Morphologies: separate droplets ($f_{\mathrm{B}}=0.50$); solvent mediated
  pillars ($f_{\mathrm{B}}=0.36$) and in-lined cylinders of B beads
  ($f_{\mathrm{B}}=0.22$).}
\end{figure}

When one moves away from $f_{\mathrm{B}}=0.5$ towards $0$, the behavior of all the
properties $I_x$, $I_z$, $G_{xx}$, $G_{zz}$ and $\sigma_{\max}^2$
differs much from the case of a one-component solvent. For the pore of
$d=20$, the bridging in $Z$ direction is observed first. This is
indicated by a large hill at $I_z$ centered around the value of
$f_{\mathrm{B}}=0.34$ (top left hand frame in figure~\ref{Fig-08r}). To be able to form
the pillar morphology, the chains need to rearrange themselves
preferentially in $Z$ direction as indicated by an increase of
$G_{zz}$ at the expense of $G_{xx}$ values (bottom left hand frame in
figure~\ref{Fig-08r}). The value of $G_{zz}=3.69$ found at $f_{\mathrm{B}}=0.34$ is
of the same order and even exceeds that observed at $f_{\mathrm{B}}=0$, namely
$G_{zz}=3.56$. This indicates that the chains within pillars are in
the regime of a good solvent. This is confirmed by the snapshots shown
in top-middle and bottom-middle columns of figure~\ref{Fig-09r}, where
the droplets merge together by the good solvent
beads. Therefore, as a result of a different miscibility of A and B
components, the solvent A nucleates by filling the gap between polymer
droplets, and the solvent mediated transition from separate droplets to
pillar morphology occurs. With a further decrease of $f_{\mathrm{B}}$, the pillars
are also bridged in $X$ direction. This is indicated by an increase of
$G_{xx}$ values and is seen in the top-right and bottom-right columns
of figure~\ref{Fig-09r}.

With an increase of the pore size $d$ to $22$, the bridging capability in
$Z$ direction is lessened giving a way for bridging along the
walls. This is demonstrated by the existence of local maxima and
minima of $I_x$, $I_z$ and $G_{xx}$ and $G_{zz}$ components (see
top-center and bottom-center frames in figure~\ref{Fig-08r}). With
a further increase of $d$ to $24$, the situation is reversed with respect
to the case of $d=20$. Now, the droplets are farther in $Z$ direction
and their bridging is observed in $X$ direction: $I_x$ and $G_{xx}$
possess maxima centered around $f_{\mathrm{B}}=0.4$ (top-right and bottom-right
frames of figure~\ref{Fig-08r}).

\begin{figure}[!t]
\centering \includegraphics[width=10cm]{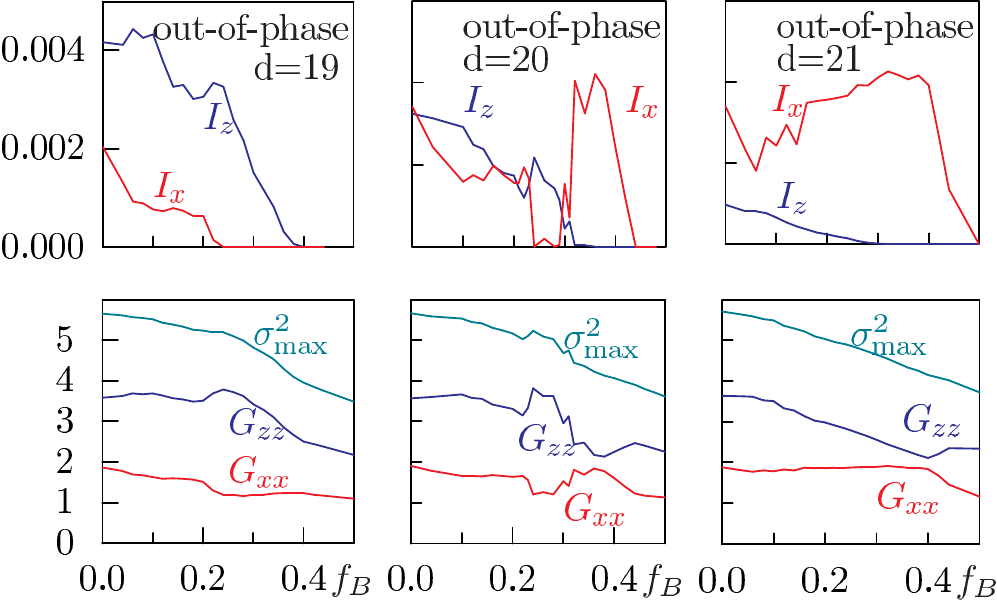}
\caption{(Color online) \label{Fig-10r}The same properties as in
figure~\protect\ref{Fig-08r} are shown for the case of out-of-phase
arrangement of stripes of the width $w=13$ at the pore
sizes $d=19, 20$ and $21$.
}
\end{figure}

Let us switch now to the case of out-of-phase arrangement of
stripes. The set of pore sizes $d=19, 20$ and $d=21$ are analyzed in
this case (the distance between brushes in $X$ and $Z$ directions is
approximately the same for $d=20$). The behavior of $I_x$, $I_z$,
$G_{xx}$, $G_{zz}$ and $\sigma_{\max}^2$ is shown in
figure~\ref{Fig-10r}. The limiting cases of $d=19$ and $d=21$ bear
similarities to their respective counterparts ($d=20$ and $d=24$) for
the in-phase arrangement of stripes. Indeed, at $d=19$, the
polymer-rich droplets are bridged in $Z$ direction first with a
decrease of $f_{\mathrm{B}}$. With a further decrease of $f_{\mathrm{B}}$, the droplets are
bridged in $X$ direction. Situation is reversed at $d=21$.

\begin{figure}[!b]
\centering \includegraphics[width=12cm]{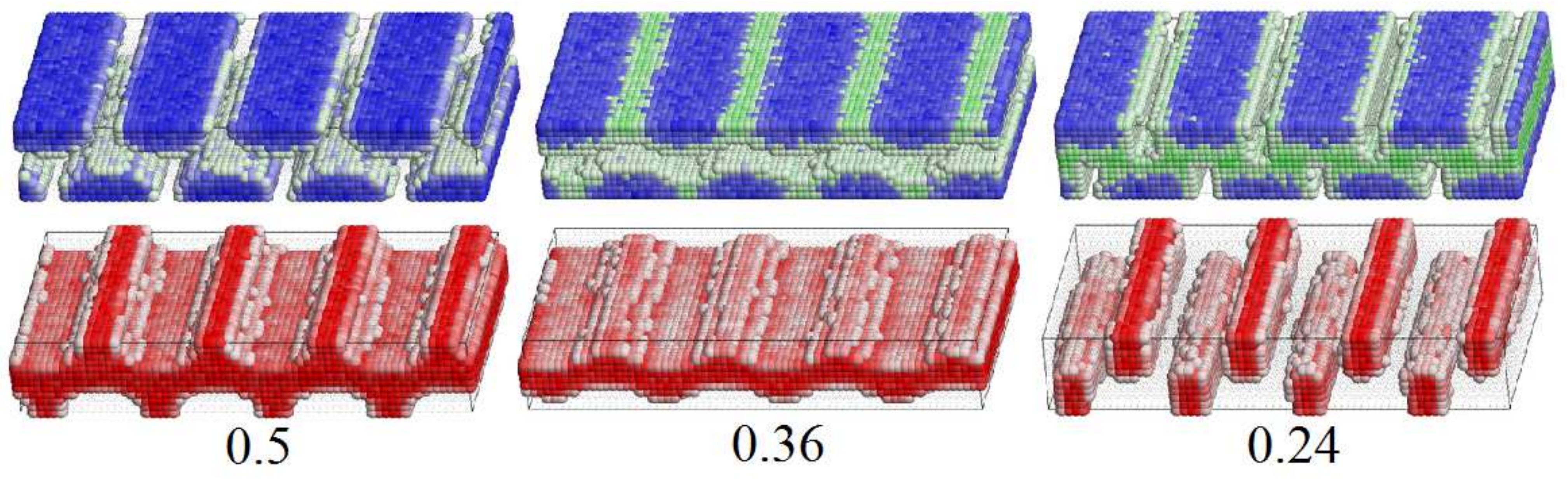}
\caption{(Color online) \label{Fig-11r}Sequence of morphologies obtained for $d=20$,
  for out-of-phase arranged stripes of the width of $w=13$ at various
  fractions of $f_{\mathrm{B}}$. Separate droplets ($f_{\mathrm{B}}=0.50$); modulated
  lamellar ($f_{\mathrm{B}}=0.36$) and meander ($f_{\mathrm{B}}=0.24$) morphologies.}
\end{figure}

The intermediate case of the pore size $d=20$ is more interesting
because it demonstrates a solvent mediated switching between various
morphologies.  When $f_{\mathrm{B}}$ decreases away from 0.5, first the bridges
in $X$ direction are formed. Both $I_x$ and $G_{xx}$ exhibit maxima
centered around $f_{\mathrm{B}}=0.36$, cf. top-middle and bottom-middle frames of
figure~\ref{Fig-10r}.  This indicates the formation of a solvent mediated
lamellar morphology, see snapshots in top-middle and bottom-middle
frames of figure~\ref{Fig-11r}. With a further decrease of $f_{\mathrm{B}}$, the
bridges are formed in $Z$ direction at the expense of those that have
been previously formed in $X$ direction.  This is indicated by a large
value of $I_z$ and by practically zero value of $I_x$ within the interval
of $f_{\mathrm{B}}\in[0.23,0.3]$. The structure of this morphology is of simple
meander (see snapshots in top-right and bottom-right hand frames of
figure~\ref{Fig-11r}).

\begin{figure}[!t]
\centering \includegraphics[width=8cm]{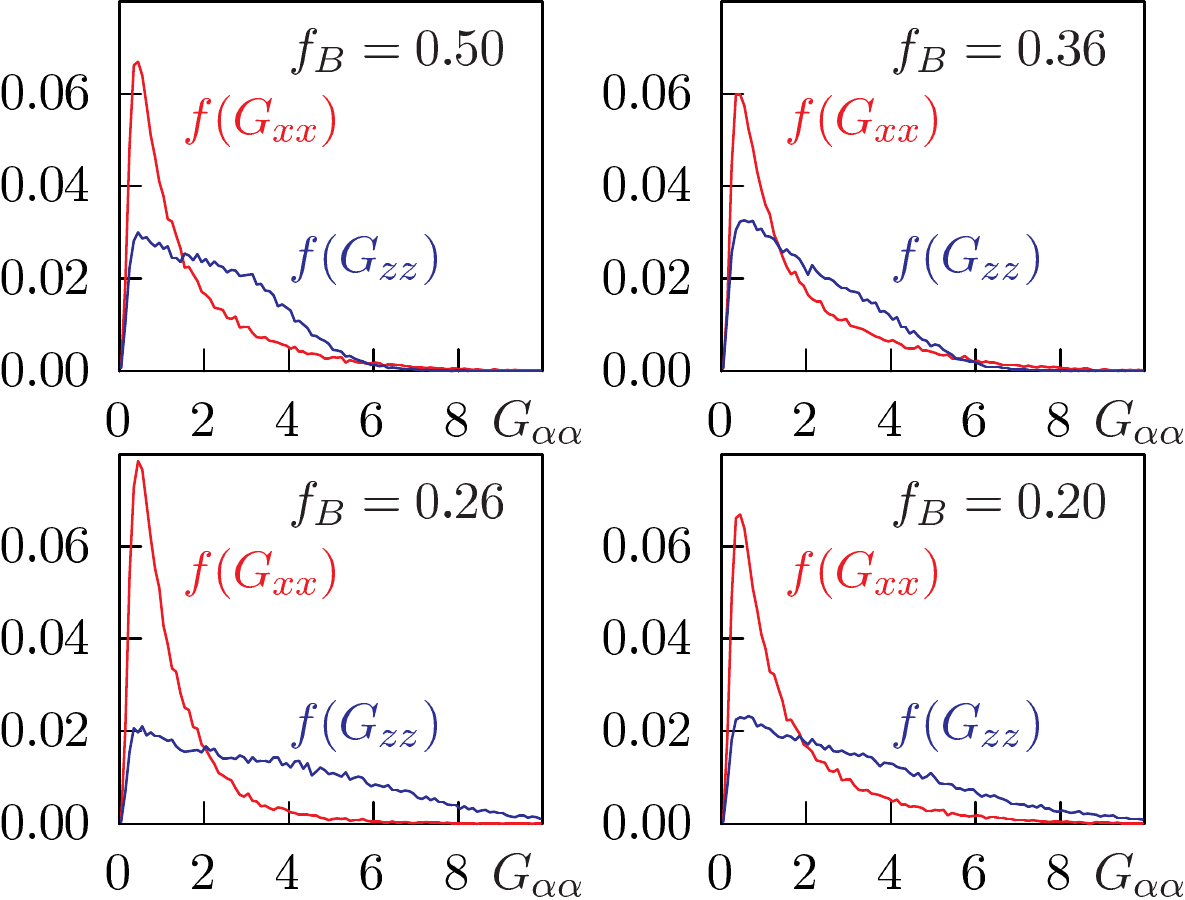}
\caption{(Color online) \label{Fig-12r}Distributions of the relevant gyration
tensor components, $G_{xx}$ and $G_{zz}$ for  various fractions $f_{\mathrm{B}}$
(indicated in each frame). The case of $d=20$, $w=13$ and out-of-phase
arrangement of stripes is shown.
}
\end{figure}

One can see that when the fraction of good solvent in a
system increases, the chains within each stripe swell and acquire certain
bistability properties. The chains can redistribute their mass either
in $X$ or in $Z$ direction (as indicated by the behavior of $G_{xx}$
and $G_{zz}$ components) and form the bridges (as indicated by
non-zero values of the overlap integrals $I_x$ and $I_z$). The
histograms for the corresponding components of the gyration tensor,
$f(G_{xx})$ and $f(G_{zz})$ provide additional insight into spatial
redistribution of the polymer chains in various morphologies. These
histograms are shown in figure~\ref{Fig-12r} for the case of pore size
$d=20$ and out-of-phase arrangement of the stripes of the width of
$w=13$.  The distribution function $f(G_{zz})$ is found to be
essentially stretched towards larger values of $G_{zz}$ for meander
($f_{\mathrm{B}}=0.26$) morphology and at still lower values of $f_{\mathrm{B}}$.  It
exhibits a quite irregular shape, whereas the distribution of
$f(G_{xx})$ is similar to the Lhuillier
form \cite{Lhuill,Victor,IlnHol} that was found as a typical
distribution of experimental radii of gyration for long
polymers. Broad distributions of a gyration tensor of both
components indicate the existence of weakly and highly deformed chains
within each stripe. It is plausible to assume that highly deformed
chains are found within the bridges that connect polymer droplets in
a solvent mediated pillar, meander and possibly other morphologies. A
measure for the average stretch of chains is provided by the maximal
eigenvalue of the gyration tensor, $\sigma_{\max}^2$. As it follows
from the set of plots shown in figures~\ref{Fig-08r} and \ref{Fig-10r},
the values of $\sigma_{\max}^2$ exhibit local maxima for certain
morphologies. The values of these maxima are higher compared to the
case of one-component solvent mapped into an effective value of
$a_{\mathrm{AC}}$ (see figure~\ref{Fig-07r}). This indicates that the loss of
conformation entropy due to an excessive average stretch of chains is
compensated by a decrease of the enthalpy caused by an increase of
contacts between similar beads of type A.

\section{Summary}\label{sec5}

In this work we consider the formation and transitions between the nanostructures
that occur inside a pore modified by stripes of tethered polymer brushes
filled with a binary mixture. The beads A of filler solvent are identical to those
of chain monomers, while the beads B exhibit a partial mixing
with beads A.
Compared to our previous study \cite{ISP1}, a number of extensions
are made. In particular, two options of in- and out-of-phase
arrangement of polymer stripes are considered. Apart from that, we
examine in detail the effect that the composition of A and B beads produces on the
formation of the morphologies in a broad range of pore geometries.
At last, for some characteristic cases we undertook supplementary
simulations with the one-component solvent of variable quality which replaces
the mixture to serve as a reference.

The change of the composition of the fluid inside the pores has a
great impact on the developing structures. For pore geometries with
narrow stripes, the problem of the description of microphase separation
inside the pore reduces to quasi two-dimensional; in the case of
moderately wide stripes and narrow pore, one faces quasi
one-dimensional demixing; whereas for very wide stripes, the system is
split into quasi two-dimensional and bulk regions.

The most interesting effects, in our view, that demonstrate solvent mediated
changes in the nanostructures occur for the geometries with weakly separated stripes
of polymer chains. With an increase of A component, the latter dissolves the polymer
chains and causes their swelling. As a result, the system acquires a
bistability in terms of either bridging the adjacent stripes along the wall or bridging the
opposite stripes across the pore. Stable morphology is formed due to competition between segregation of A and B beads and the deformation of the
chains. In our simulations we observe morphology switching due to subtle changes
in pore geometry and/or in the composition of a mixture.

We found the following solvent mediated morphologies: in-lined cylinders (made of one
component), meander structure and wave-shaped modulated internal channels.
The suggested applications of such structures (possibly, after making the
structure permanent via crosslinking) involve nanopatterning for
manufacturing of nanochannels, nanorods and similarly sized objects.

So far our studies concentrated on the equilibrium DPD
simulations. However, it would be of interest to check how the
morphologies change if the fluid undergoes a pressure-driven flow
along the pore axis. This problem is currently under study in our
laboratories.

\section*{Acknowledgements}

This work was supported by EC under the Grant No. PIRSES~268498.
J.~I. is thankful to T.~Kreer, S.~Santer and D.~Neher for fruitful and
stimulating discussions.

\newpage

\ukrainianpart

\title{Фазові переходи за посередництва розчинника у порах \\ із стінками декорованими полімерними
  щітками. Комп'ютерне моделювання методом \\ дисипативної динаміки}
\author{Я.М. Ільницький\refaddr{a1}, С.Соколовскі\refaddr{a2},
  Т. Пацаган\refaddr{a1}}
\addresses{
\addr{label1} Інститут фізики конденсованих систем НАН України,
вул. І. Свєнціцького, 1, 79011 Львів, Україна
\addr{label2} Відділ моделювання фізико-хіиічних процесів, Університет
Марії Кюрі-Склодовської, \\ 20031 Люблін, Польща }

\makeukrtitle

\begin{abstract}
\tolerance=3000%

Розглянуто процес самоформування фаз у порі із стінками декорованими
полосками полімерної щітки (з мономерів сорту А), яка заповнена
бінарною сумішшю із компонент А і В. Акцент зроблено на дослідженні
ролі розчинника при фазових переходах спричинених зміною складу
суміші. Знайдено граничні випадки квазиодновимірного та
квазидвовимірного незмішування. Показано, що механізм формування
морфологій (та, в деяких випадках, переключення між ними) при зміні
складу суміші полягає у зміні локального середовища для ланцюжків
полімерних щіток. Знайдено сформовані за посередництва розчинника
ламеларні, меандро-подібні та порядкові циліндричні фази. Структури
вивчено за допомогою аналізу інтегралів перекриття щіток та величин
компонент тензора гірації.

\keywords дисипативна динаміка, суміші, пори, наноструктури
\end{abstract}

\begin{thebibliography}{99}
\bibitem{1}  De Gennes P.G.,  Rev. Mod. Phys., 1985, \textbf{57}, 827; \doi{10.1103/RevModPhys.57.827}.

\bibitem{2}  Advincula R.C.,  Brittain W.J.,  Caster K.C., Ruhe J., {Polymer Brushes:
Synthesis, Characterization, Applications}, Willey, New York, 2004.

\bibitem{3}   R\"uhe J.,   Ballauf M.,   Biesalski M.,   Dziezok P.,
 Grohn F.,  Johannsmann D.,  Houbenov N.,  Hugenberg N.,   Konradi R.,
 Minko S.,  Motornov M.,  Netz R.R.,  Schmidt M.,  Seidel C.,  Stamm M.,
 Stephan T.,  Usov D.,  Zhang H.N., Adv. Polym. Sci., 2004, \textbf{165}, 79; \doi{10.1007/b11268}.

\bibitem{4}   Naji A.,  Seidel C.,  Netz R.R., Adv. Polym. Sci., 2006, \textbf{198}, 149; \doi{10.1007/12_062}.

\bibitem{5}  Klushin L.I.,  Skvortsov A.M., J. Phys. A: Math. Theor., 2011, \textbf{44}, 473001; \doi{10.1088/1751-8113/44/47/473001}.

\bibitem{6}  Binder K.,  Kreerb T.,  Milchev A., Soft Matter, 2011, \textbf{7},  7159; \doi{10.1039/c1sm05212h}.

\bibitem{7}  {Polymer Brushes}, Advincula R.C.,  Brittain W.J.,  Caster K.C.,  R\"uhe J. (Eds.), Wiley-VCH, Weinheim, 2004.

\bibitem{9}  Descas R.,  Sommer J.-U.,   Blumen A., Macromol. Theory Simul., 2008, \textbf{17},
429; \doi{10.1002/mats.200800046}.

\bibitem{a1}  Garbassi F.,  Morra M.,  Occhiello E., { Polymer Surfaces: From Physics to
Technology}, John Wiley \& Sons;  New York, 2002.

\bibitem{a2}  Sperling L.H., {Polymeric Multicomponent Materials: An Introduction
LED}, Wiley Interscience, New York, 1997.

\bibitem{a3}  Klein J., Science, 2009, \textbf{323}, 47; \doi{10.1126/science.1166753}.

\bibitem{a4}  Napper D.H., {Polymeric Stabilization of Colloidal Dispersions},
Academic, London, 1983.

\bibitem{a5}  Storm G.,  Belliot S.O.,  Daemen T.,  Lasic D.D., Adv. Drug
Delivery Rev., 1995, \textbf{17}, 31; \\ \doi{10.1016/0169-409X(95)00039-A}.

\bibitem{a6}  Hucknall A.,  Rangarajan S.,  Chilkoti A.,
Adv. Mater., 2009, \textbf{21},    2441; \doi{10.1002/adma.200900383}.

\bibitem{a7}  Wang A.J.,  Xu J.J.,  Chen H.Y., J. Chromatogr. A, 2007, \textbf{1147}, 120; \doi{10.1016/j.chroma.2007.02.030}.

\bibitem{a8}    Li Y.,  Zhang J.,  Fang L.,  Wang T.,  Zhu S.,  Li Y.,  Wang Z.,   Zhang L.,
 Cui L.,  Yang B., Small, 2011, \textbf{7}, 2769; \\ \doi{10.1002/smll.201100313}.

\bibitem{r1}  Guskova O.A.,  Seidel C., Macromolecules, 2011,  \textbf{44}, 671; \doi{10.1021/ma102349k}.

\bibitem{r2}   Wang J.,  M\"uller M.,  Macromolecules, 2009,  \textbf{42} , 2251; \doi{10.1021/ma8026047}.

\bibitem{r3}  Yin Y.,   Sun P.,  Li B.,  Chen T.,   Jin Q.,   Ding D.,  Shi A.-C.,
Macromolecules, 2007, \textbf{40}, 5161; \doi{10.1021/ma070393n}.

\bibitem{r4}  Bormashenko E.,  Pogreb R.,  Stanevsky O.,   Bormashenko Y.,  Tamir S.,
 Cohen R.,  Nunberg M.,   Gaisin V.Z.,  Gorelik M.,  Gendelman O.V.,
Mater. Lett., 2005,  \textbf{59}, 2461; \doi{10.1016/j.matlet.2005.03.015}.

\bibitem{r5}  Nagpal U.,  Kang H.,  Craig G.S.W.,  Nealey P.F.,  de Pablo J.J.,
ACS Nano,  2011, \textbf{5}, 5673; \doi{10.1021/nn201335v}.

\bibitem{r6}  Cao Q.,   Zuo C.,  Li L.,
 Yang Y.,   Li N., Microfluid Nanofluid, 2011,  \textbf{10}, 977; \doi{10.1007/s10404-010-0726-9}.

\bibitem{nt}  Chen T.,  Amin I.,  Jordan R.,   Chem. Soc. Rev., 2012,
\textbf{41}, 3280; \doi{10.1039/c2cs15225h}.

\bibitem{th1}  Guangsuo L., {Fabrication of nanopatterns via surface chemical modification and
reactive reversal nanoimprint lithography}, PhD Thesis, National University of Singapore,
Singapore, 2010, \\ \url{http://scholarbank.nus.edu.sg/handle/10635/22850}.

\bibitem{lito}  Tolfree D.W.L., Rep. Prog. Phys., 1998, \textbf{61}, 313; \doi{10.1088/0034-4885/61/4/001}.
%
\bibitem{lito_1} Burmeister F.,  Schl\"afle C.,  Keilhofer B.,  Bechinger C.,  Boneberg J.,
Leiderer P., Adv. Mater., 1998, \textbf{10}, 495; \doi{10.1002/(SICI)1521-4095(199804)10:6<495::AID-ADMA495>3.0.CO;2-A}.
%
\bibitem{lito_2} Knight J.B.,  Vishwanath A.,  Brody J.P.,  Austin R.H., Phys. Rev. Lett., 1998, \textbf{80}, 3863; \\ \doi{10.1103/PhysRevLett.80.3863}.

\bibitem{nano} {Surface-Initiated Polymerization I and II}, Jordan R. (Ed.),
Adv. Polym. Sci.,  \textbf{197}-\textbf{198}, 2006.
%
\bibitem{nano_1} Zhou X.,  Chen Y.,  Li B.,  Lu G.,  Boey F.Y.C.,  Ma J.,  Zhang H.,
Small, 2008, \textbf{4}, 1324; \doi{10.1002/smll.200701267}.
%
\bibitem{nano_2} Schmelmer U.,  Paul A.,  K\"uller A.,  Steenackers M.,
 Ulman A.,  Grunze M.,  G\"olzh\"auser A.,  Jordan R.,
Small, 2007, \textbf{3}, 459; \doi{10.1002/smll.200600528}.

\bibitem{mc}
Adamczyk P., Romiszowski P., Sikorski A., Catal. Lett., 2009, \textbf{129}, 130; \doi{10.1007/s10562-008-9795-8}.
%
\bibitem{mc_1} Chang C.-Y.,  Yang H.-W.,  Lin J.-S.,  Ju S.-P.,  Hsieh J.-Y., J. Comput. Theor. Nanosci., 2011, \textbf{8},2439; \\  \doi{10.1166/jctn.2011.1976}.
%
\bibitem{mc_2} Singh S.K.,  Khan S.,  Jana S.,  Singh J.K.,
Mol. Simulat., 2011, \textbf{37}, 350; \doi{10.1080/08927022.2010.514778}.
%
\bibitem{mc_3} Koutsioubas A.G.,  Vanakaras A.G., Langmuir, 2008, \textbf{24}, 13717; \doi{10.1021/la802536v}.
%
\bibitem{mc_4} Chen H.,  Chen X.,  Ye Z.,  Liu H.,  Hu Y., Langmuir, 2010, \textbf{26}, 6663; \doi{10.1021/la904001h}.

\bibitem{md}  Jayaraman A.,  Hall C.K.,  Genzer J., J. Chem. Phys., 2005, \textbf{123}, 124702; \doi{10.1063/1.2043048}.
%
\bibitem{md_1} Chen H.,  Peng C.,  Sun L., Liu H.,  Hu Y.,  Jiang J., Langmuir, 2007, \textbf{23}, 11112; \doi{10.1021/la701773a}.

\bibitem{dpd}   Spaeth J.R.,  Dale T.,  Kevrekidis I.G.,
Panagiotopoulos A.Z.,  Ind. Eng. Chem. Res., 2011, \textbf{50}, 69; \\ \doi{10.1021/ie100337r}.
%
\bibitem{dpd_1} Patra M.,  Linse P., Nano Lett., 2006, \textbf{6}, 133; \doi{10.1021/nl051611y}.
%
\bibitem{dpd_2} Goujon F.,  Malfreyt P.,  Tildesley D.J., J. Chem. Phys., 2008,
\textbf{129}, 034902; \doi{10.1063/1.2954022}.
%
\bibitem{dpd_3} Li C.-S.,  Wu W.-C.,  Sheng Y.-J.,  Chen W.-C., J. Chem. Phys., 2008,
\textbf{128}, 154908; \doi{10.1063/1.2904866}.
%
\bibitem{dpd_4} Petrus P., L\'isal M.,  Brennan J.K., Langmuir, 2010,
\textbf{26} 3695; \doi{10.1021/la903200j}.
%
\bibitem{dpd_5} Petrus P., L\'isal M.,  Brennan J.K., Langmuir, 2010,
\textbf{26}, 14680; \doi{10.1021/la102666g}.

\bibitem{ISP1}  Ilnytskyi J.M.,  Patsahan T.,  Soko\l owski S.,
J. Chem. Phys., 2011, \textbf{134},  204903; \doi{10.1063/1.3592562}.

\bibitem{IL}  Ilnytskyi J.M.,  Patsahan T.,  Holovko M.,  Krouskop P.E.,
Makowski M.P., Macromolecules, 2008, \textbf{41}, 9904; \\ \doi{10.1021/ma801045z}.

\bibitem{GrWarr}
 Groot R.D.,  Warren P.B.,  J. Chem. Phys., 1997, \textbf{107}, 4423; \doi{10.1063/1.474784}.

\bibitem{Zhao}
Zhao Y.,  Liu H.,  Lu Zh.,  Sun Ch., Chin. J. Chem. Phys., 2008, \textbf{21}, 451; \doi{10.1088/1674-0068/21/05/451-456}.

\bibitem{Romeis_2012}
Romeis D., Merlitz  H.,  Sommer J.U., J. Chem. Phys., 2012, \textbf{136}, 044903; \doi{10.1063/1.3676657}.

\bibitem{deGennes}
De Gennes P.G., Scaling Concepts in Polymer Physics, Cornell University Press, Ithaca and London,
1979.

\bibitem{IlnHol}  Ilnytskyi J.M.,  Holovatch Yu.,
Condens. Matter Phys., 2007, \textbf{10}, 539; \doi{10.5488/CMP.10.4.539}.

\bibitem{Kreer_2001}
 Kreer T.,  M\"{u}ser M.H.,  Binder K.,  Klein J., Langmuir, 2001, \textbf{17}, 7804; \doi{10.1021/la010807k}.

\bibitem{Lhuill} Lhuillier D., J. Phys., 1988, \textbf{49}, 705; \doi{10.1051/jphys:01988004905070500}.

\bibitem{Victor} Victor J.M., Lhuillier D., J. Chem. Phys., 1990, \textbf{92},
1362; \doi{10.1063/1.458147}.

\end{thebibliography}
\end{document}